\documentclass[conference,10pt,twocolumn,a4paper]{IEEEtran}

\usepackage[T1]{fontenc}
\usepackage[applemac]{inputenc}

\usepackage{graphicx}
\usepackage[hang]{subfigure}
\usepackage{tikz}
\usepackage{caption}
\usepackage{pgfplots}

\usepackage[sort]{cite}

\usepackage{amssymb}
\usepackage{amsmath}
\usepackage{paralist}

\usepackage{functan}
\usepackage{url}

\newtheorem{thm}{Theorem}
\newtheorem{cor}[thm]{Corollary}   

\newtheorem{defn}{Definition}

\newtheorem{lem}[thm]{Lemma}

\DeclareMathOperator{\st}{subject\ to}
\newcommand{\supp}{\text{supp}}

\DeclareMathOperator*{\minimize}{minimize}

\newcommand{\sigmaMin}{\sigma_\text{min}}

\newcommand{\minNorm}{\omega_\text{min}}
\newcommand{\maxNorm}{\omega_\text{max}}
\newcommand{\maxCorr}{\mu}
\newcommand{\maxCorrFirst}{\mu_1}
\newcommand{\maxCorrSecond}{\mu_2}
\newcommand{\mutualMaxCorr}{\mu_m}
\newcommand{\coh}{\hat{\mu}}
\newcommand{\cohFirst}{\hat{\mu}_1}
\newcommand{\cohSecond}{\hat{\mu}_2}
\newcommand{\mutualCoh}{\hat{\mu}_m}

\newcommand{\C}{\mathbb{C}}
\newcommand{\R}{\mathbb{R}}

\newcommand{\Z}{\mathbb{Z}}
\newcommand{\suppSig}{\mathcal{S}}

\newcommand{\maxCoeffSet}{\mathcal{Q}}
\newcommand{\maxCoeffSetFirst}{\mathcal{Q}_1}
\newcommand{\maxCoeffSetSecond}{\mathcal{Q}_2}

\newcommand{\analysis}{\boldsymbol{\Psi}}
\newcommand{\analysisFirst}{\boldsymbol{\Psi}_1}
\newcommand{\analysisSecond}{\boldsymbol{\Psi}_2}
\newcommand{\analysisCompound}{\boldsymbol{\Psi}}
\newcommand{\measMat}{\mathbf{A}}
\newcommand{\measMatFirst}{\mathbf{A}_1}
\newcommand{\measMatSecond}{\mathbf{A}_2}
\newcommand{\measMatCompound}{\mathbf{A}}
\newcommand{\proj}{\mathbf{P}}
\newcommand{\zeroMat}{\mathbf{0}}
\newcommand{\idMat}{\mathbf{I}}
\newcommand{\dico}{\mathbf{D}}
\newcommand{\dicoFirst}{\mathbf{D}_1}
\newcommand{\dicoSecond}{\mathbf{D}_2}

\newcommand{\meas}{\mathbf{y}}
\newcommand{\sig}{\mathbf{x}}
\newcommand{\sigOpt}{\mathbf{x}^*}
\newcommand{\sigProb}{\widetilde{\sig}}
\newcommand{\sigFirst}{\mathbf{x}_1}
\newcommand{\sigFirstOpt}{\mathbf{x}_1^*}
\newcommand{\sigFirstProb}{\widetilde{\mathbf{x}}_1}
\newcommand{\sigSecond}{\mathbf{x}_2}
\newcommand{\sigSecondOpt}{\mathbf{x}_2^*}
\newcommand{\sigSecondProb}{\widetilde{\mathbf{x}}_2}
\newcommand{\sigCompound}{\mathbf{x}}
\newcommand{\sigCompoundOpt}{\mathbf{x}^*}
\newcommand{\sigCompoundProb}{\widetilde{\mathbf{x}}}
\newcommand{\sparseSig}{\mathbf{s}}

\newcommand{\sparseSigFirstProb}{\mathbf{\tilde{s}}_1}

\newcommand{\sparseSigSecondProb}{\mathbf{\tilde{s}}_2}
\newcommand{\noise}{\mathbf{e}}
\newcommand{\kerVec}{\mathbf{h}}
\newcommand{\kerVecCompound}{\mathbf{h}}

\newcommand{\dimSig}{d}

\newcommand{\dimSigFirstHybrid}{p_1}
\newcommand{\dimSigSecondHybrid}{p_2}
\newcommand{\dimSigHybrid}{p}
\newcommand{\dicoAtomNb}{n}

\newcommand{\sizeAnalysis}{n}
\newcommand{\sizeAnalysisFirst}{n_1}
\newcommand{\sizeAnalysisSecond}{n_2}
\newcommand{\sparsity}{k}
\newcommand{\sparsityFirst}{k_1}
\newcommand{\sparsitySecond}{k_2}
\newcommand{\measNb}{m}

\newcommand{\colIdx}{j}
\newcommand{\rowIdx}{i}

\newcommand{\prob}{\text{P}^*}
\newcommand{\probSplit}{\text{P}}
\newcommand{\probSplitSynthesis}{\text{PS}}
\newcommand{\probSplitAnalysis}{\text{PA}}

\newcommand{\thetaMin}{\theta_\text{min}}
\newcommand{\thetaMax}{\theta_\text{max}}

\begin{document}

\IEEEoverridecommandlockouts 

\title{Sparse Signal Separation in Redundant Dictionaries}

\author{
	\IEEEauthorblockN{C\'eline Aubel\IEEEauthorrefmark{1}, Christoph Studer\IEEEauthorrefmark{2}, Graeme Pope\IEEEauthorrefmark{1}, and Helmut B\"olcskei\IEEEauthorrefmark{1}}\\[-0.0cm]
	\IEEEauthorblockA{\IEEEauthorrefmark{1}Dept.~of IT \& EE, ETH Zurich, 8092 Zurich, Switzerland \\
	\IEEEauthorrefmark{2}Dept.~of ECE, Rice University, Houston, TX, USA\\
	Email: \{aubelc, gpope, boelcskei\}@nari.ee.ethz.ch, studer@rice.edu}
	\vspace{-0.3cm}
	\thanks{The work of C.~Studer was supported by the Swiss National Science Foundation (SNSF) under Grant PA00P2-134155.}
}

\maketitle


\begin{abstract}

We formulate a unified framework for the separation of signals that are sparse in ``morphologically'' different redundant dictionaries. This formulation incorporates the so-called ``analysis'' and ``synthesis''  approaches as special cases and contains novel hybrid setups. We find corresponding coherence-based recovery guarantees for an $\ell_1$-norm based separation algorithm. Our results recover those reported in Studer and Baraniuk, ACHA, \textit{submitted}, for the synthesis setting, provide new recovery guarantees for the analysis setting, and form a basis for comparing performance in the analysis and synthesis settings.  As an aside our findings complement the D-RIP recovery results reported in Cand\`es \textit{et al.}, ACHA, 2011, for the ``analysis'' signal recovery problem 
$$\underset{\sigProb}{\text{minimize}} \,\, \|\analysis\sigProb\|_1 \quad \text{subject to} \,\, \|\meas - \measMat\sigProb\|_2 \leq \varepsilon$$
by delivering corresponding coherence-based recovery results.

\end{abstract}

\section{Introduction}
\label{intro}

We consider the problem of splitting the signal $\sig = \sigFirst + \sigSecond$ into its constituents $\sigFirst \in \C^\dimSig$ and $\sigSecond \in \C^\dimSig$---assumed\linebreak[4]to be sparse in ``morphologically'' different (redundant) dictionaries~\cite{Rubinstein:2010p186}---based on $\measNb$ linear, nonadaptive, and noisy measurements $\meas = \measMat\sig + \noise$.
Here, $\measMat \in \C^{\measNb \times \dimSig}$, $\measNb \leq \dimSig$, is the measurement matrix, assumed to be known, and $\noise \in \C^\measNb$ is a noise vector, assumed to be unknown and to satisfy $\|\noise\|_2 \leq \varepsilon$, with $\varepsilon$ known.

Redundant dictionaries~\cite{frames, DBLP:books/daglib/0025129} often lead to sparser representations than nonredundant ones, such as, e.g., orthonormal bases, and have therefore become pervasive in the sparse signal recovery literature~\cite{DBLP:books/daglib/0025129}. In the context of signal separation, redundant dictionaries lead to an interesting dichotomy~\cite{Milanfar:2007p708, Rubinstein:2010p186, Candes:2011p12}:
\begin{itemize}
 	\item In the so-called ``synthesis'' setting, it is assumed that, for $\ell = 1, 2$, $\sig_\ell = \dico_\ell\sparseSig_\ell$, where $\dico_\ell \in \C^{\dimSig \times \dicoAtomNb}$ ($\dimSig < \dicoAtomNb$) is a redundant dictionary (of full rank) and the coefficient vector $\sparseSig_\ell \in \C^\dicoAtomNb$ is sparse (or approximately sparse in the sense of~\cite{Candes:2005p164}). 
Given the vector $\meas \in \C^\measNb$, the problem of finding the constituents $\sigFirst$ and $\sigSecond$ is formalized as~\cite{Studer:2012p15}:
\begin{equation}
	(\probSplitSynthesis) \left\{\begin{array}{ll}
						\displaystyle\minimize_{\sparseSigFirstProb, \sparseSigSecondProb}&\|\sparseSigFirstProb\|_1 + \|\sparseSigSecondProb\|_1 \\
						\st & \|\meas - \measMat(\dicoFirst\sparseSigFirstProb + \dicoSecond\sparseSigSecondProb)\|_2 \leq \varepsilon.
				      \end{array}\right.\notag
\end{equation}

	\item In the so-called ``analysis'' setting, it is assumed that, for $\ell = 1, 2$, there exists a matrix $\analysis_\ell\in \C^{\sizeAnalysis \times \dimSig}$ such that $\analysis_\ell\sig_\ell$ is sparse (or approximately sparse). 
The problem of recovering $\sigFirst$ and $\sigSecond$ from $\meas$ is formalized as~\cite{Candes:2011p12}:
\begin{equation}
	(\probSplitAnalysis) \left\{\begin{array}{ll}
						\underset{\sigFirstProb, \sigSecondProb}{\minimize}&\|\analysisFirst\sigFirstProb\|_1 + \|\analysisSecond\sigSecondProb\|_1 \\
						\st &\|\meas - \measMat(\sigFirstProb + \sigSecondProb)\|_2 \leq \varepsilon.
				      \end{array}\right.\notag
\end{equation}

\end{itemize}
Throughout the paper, we exclusively consider redundant dictionaries as for $\dico_\ell$, $\ell = 1, 2$, square, the synthesis setting can be recovered from the analysis setting by taking $\analysis_\ell = \dico_\ell^{-1}$.

The problems $(\probSplitSynthesis)$ and $(\probSplitAnalysis)$ arise in numerous applications including denoising~\cite{Mallat_WaveletTourOfSignalProcessing}, super-resolution~\cite{Mallat_WaveletTourOfSignalProcessing}, inpainting~\cite{Elad:2005p947,Fadili:2010p977, Cai:2010p4236}, deblurring~\cite{Cai:2010p4236}, and recovery of sparsely corrupted signals~\cite{Studer:2012p4592}. Coherence-based recovery guarantees for $(\probSplitSynthesis)$ were reported in~\cite{Studer:2012p15}. The problem~$(\probSplitAnalysis)$ was mentioned in~\cite{Candes:2011p12}. In the noiseless case, recovery guarantees for $(\probSplitAnalysis)$, expressed in terms of a concentration inequality, are given in~\cite{kutyniok} for $\measMat = \idMat_\dimSig$ and $\analysisFirst$ and $\analysisSecond$ both Parseval frames~\cite{frames}.

\subsubsection*{Contributions} We consider the general problem 
\begin{equation}
	(\probSplit) \left\{\begin{array}{ll}
						\underset{\sigFirstProb, \sigSecondProb}{\minimize}&\|\analysisFirst\sigFirstProb\|_1 + \|\analysisSecond\sigSecondProb\|_1 \\
						\st & \|\meas - \measMatFirst\sigFirstProb - \measMatSecond\sigSecondProb\|_2 \leq \varepsilon,
				      \end{array}\right.\notag
\end{equation}
which encompasses $(\probSplitSynthesis)$ and $(\probSplitAnalysis)$.
To recover $(\probSplitSynthesis)$ from ($\probSplit$), one sets $\measMat_\ell = \measMat\dico_\ell$ and $\analysis_\ell = [\idMat_\dimSig\ \zeroMat_{\dimSig, \sizeAnalysis-\dimSig}]^T$, for $\ell = 1, 2$. $(\probSplitAnalysis)$~is obtained by choosing
$\measMat_\ell = \measMat$, for $\ell = 1, 2$.
Our main contribution is a coherence-based recovery guarantee for the general problem~$(\probSplit)$. This result recovers~\cite[Th.~4]{Studer:2012p15}, which deals with $(\probSplitSynthesis)$, provides new recovery guarantees for~$(\probSplitAnalysis)$, and constitutes a basis for comparing performance in the analysis and synthesis settings. As an aside, it also complements the D-RIP recovery guarantee in~\cite[Th.~1.2]{Candes:2011p12} for the problem
\begin{equation}
	(\prob)\,\,\, \underset{\sigProb}{\minimize}\,\,\|\analysis\sigProb\|_1 \quad \st\,\,\|\meas - \measMat\sigProb\|_2 \leq \varepsilon \notag
\end{equation}
by delivering a corresponding coherence-based recovery guarantee.
Moreover, the general formulation $(\probSplit)$ encompasses novel hybrid problems of the form
\begin{equation}
	\begin{array}{ll}
		\underset{\sparseSigFirstProb, \sigSecondProb}{\minimize}&\|\sparseSigFirstProb\|_1 + \|\analysisSecond\sigSecondProb\|_1 \\
		\st & \|\meas - \measMat(\dicoFirst\sparseSigFirstProb - \sigSecondProb)\|_2 \leq \varepsilon.
	\end{array}\notag
\end{equation}


\setlength\fboxsep{0pt}    
 \setlength\fboxrule{0.1pt}
\begin{figure*}
	\center
	\subfigure[Original cartoon image]{\fbox{\includegraphics[width = 0.32\textwidth]{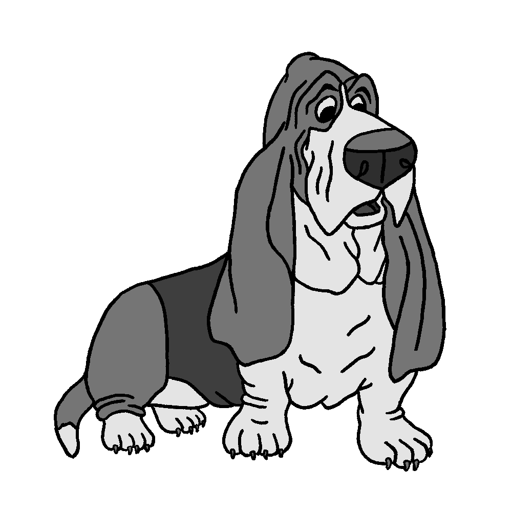}}\label{cartoon}}
	~\subfigure[Corrupted image]{\fbox{\includegraphics[width = 0.32\textwidth]{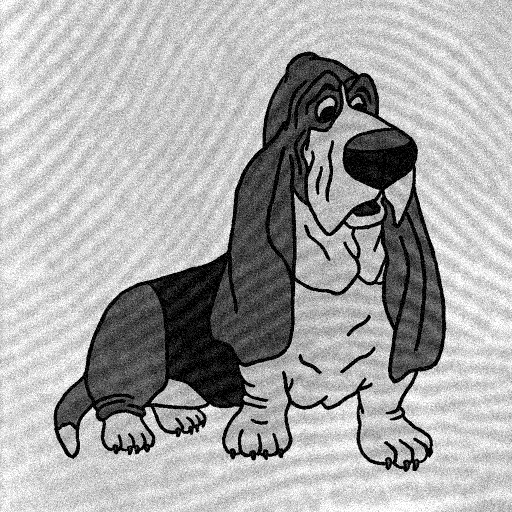}}\label{input}}
	~\subfigure[Restored cartoon image]{\fbox{\includegraphics[width = 0.32\textwidth]{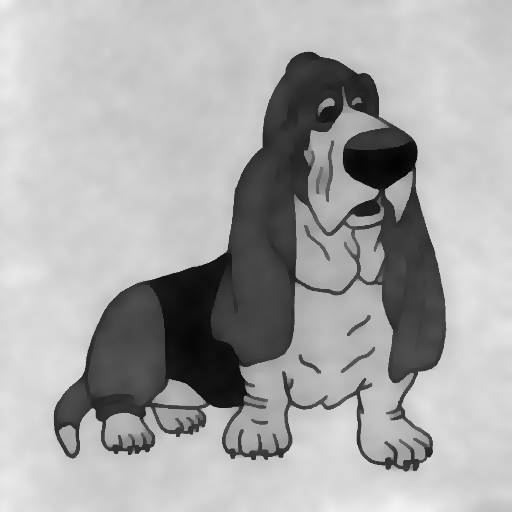}}\label{cartoon restored}}
	\caption{Image separation in the presence of Gaussian noise ($\text{SNR} = 20\,\text{dB}$).}
	\label{with noise}
\end{figure*}

\subsubsection*{Notation}

Lowercase boldface letters stand for column vectors and uppercase boldface letters denote matrices. The transpose, conjugate transpose, and Moore-Penrose inverse of the matrix $\mathbf{M}$ are designated as $\mathbf{M}^T$, $\mathbf{M}^H$, and $\mathbf{M}^\dagger$, respectively. The $\colIdx$th column of $\mathbf{M}$ is written~$[\mathbf{M}]_\colIdx$, and the entry in the $\rowIdx$th row and $\colIdx$th column of $\mathbf{M}$ is $[\mathbf{M}]_{\rowIdx, \colIdx}$. We let $\sigmaMin(\mathbf{M})$ denote the smallest singular value of~$\mathbf{M}$, use $\idMat_n$ for the $n \times n$ identity matrix, and let $\zeroMat_{k \times m}$ be the $k \times m$ all zeros matrix. For matrices~$\mathbf{M}$ and $\mathbf{N}$, we let $\minNorm(\mathbf{M}) \triangleq \min_\colIdx\|[\mathbf{M}]_j\|_2$, $\maxNorm(\mathbf{M}) \triangleq \max_\colIdx\|[\mathbf{M}]_j\|_2$, $\minNorm(\mathbf{M}, \mathbf{N}) \triangleq \min\{\minNorm(\mathbf{M}), \minNorm(\mathbf{N})\}$, and $\maxNorm(\mathbf{M}, \mathbf{N}) \triangleq \max\{\maxNorm(\mathbf{M}), \maxNorm(\mathbf{N})\}$.
The $k$th entry of the vector~$\mathbf{x}$ is written~$[\mathbf{x}]_k$, and $\|\mathbf{x}\|_1 \triangleq \sum_k|[\mathbf{x}]_k|$ stands for its $\ell_1$-norm. We take $\supp_k(\mathbf{x})$ to be the set of indices corresponding to the $k$ largest (in magnitude) coefficients of $\mathbf{x}$. 
Sets are designated by uppercase calligraphic letters; the cardinality of the set~$\mathcal{S}$ is $|\mathcal{S}|$ and the complement of $\mathcal{S}$ (in some given set) is denoted by $\mathcal{S}^c$.  For a set $\mathcal{S}$ of integers and $n \in \Z$, we let $n+ \mathcal{S} \triangleq \{n+p : p \in \mathcal{S}\}$.
The $n \times n$ diagonal projection matrix $\proj_\mathcal{S}$ for the set $\mathcal{S} \subset \{1, \ldots, n\}$ is defined as follows:
\begin{equation}
	[\proj_\mathcal{S}]_{\rowIdx, \colIdx} = \left\{\begin{array}{ll}
								1, & \rowIdx = \colIdx\text{ and }\rowIdx \in \mathcal{S}\\
								0, & \text{otherwise,}
							\end{array}\right.\notag
\end{equation}
and we set $\mathbf{M}_\mathcal{S} \triangleq \proj_\mathcal{S}\mathbf{M}$.
We define $\sigma_k(\mathbf{x})$ to be the $\ell_1$-norm approximation error of the best $k$-sparse approximation of $\mathbf{x}$, i.e., $\sigma_k(\mathbf{x}) \triangleq \|\mathbf{x} - \mathbf{x}_\mathcal{S}\|_1$ where $\mathcal{S} = \supp_k(\mathbf{x})$ and $\mathbf{x}_\mathcal{S} \triangleq \mathbf{P}_\mathcal{S}\mathbf{x}$.

\section{Recovery Guarantees}
\label{mainResult}

Coherence definitions in the sparse signal recovery literature~\cite{DBLP:books/daglib/0025129} usually apply to dictionaries with normalized columns. Here, we will need coherence notions valid for general (unnormalized) dictionaries $\mathbf{M}$ and $\mathbf{N}$, assumed, for simplicity of exposition, to consist of nonzero columns only.

\begin{defn}[Coherence]
	The coherence of the dictionary~$\mathbf{M}$ is defined as
	\begin{equation}
		\coh(\mathbf{M}) = \underset{\rowIdx, \colIdx, \rowIdx \neq \colIdx}{\max}\,\,\frac{|[\mathbf{M}^H\mathbf{M}]_{\rowIdx,\colIdx}|}{\minNorm^2(\mathbf{M})}.
	\end{equation}
	\label{def coherence}
\end{defn}

\begin{defn}[Mutual coherence]
	The mutual coherence of the dictionaries~$\mathbf{M}$ and $\mathbf{N}$ is defined as
	\begin{equation}
		\coh_m(\mathbf{M}, \mathbf{N}) = \underset{\rowIdx, \colIdx}{\max}\,\,\frac{|[\mathbf{M}^H\mathbf{N}]_{\rowIdx,\colIdx}|}{\minNorm^2(\mathbf{M}, \mathbf{N})}.
	\end{equation}
	\label{def mutual coherence}
\end{defn}

\vspace{-0.4cm}

The main contribution of this paper is the following recovery guarantee for $(\probSplit)$.

\begin{thm}
	Let $\meas = \measMatFirst\sigFirst + \measMatSecond\sigSecond + \noise$ with $\|\noise\|_2 \leq \varepsilon$ and let $\analysisFirst \in \C^{\sizeAnalysisFirst \times \dimSigFirstHybrid}$ and $\analysisSecond \in \C^{\sizeAnalysisSecond \times \dimSigSecondHybrid}$ be full-rank matrices. Let $\sig = [\sigFirst^T \ \sigSecond^T]^T$, $\cohFirst = \coh(\measMatFirst\analysisFirst^\dagger)$, $\cohSecond = \coh(\measMatSecond\analysisSecond^\dagger)$, $\mutualCoh = \coh_m(\measMatFirst\analysisFirst^\dagger, \measMatSecond\analysisSecond^\dagger)$, and $\coh_\text{max} = \max\{\cohFirst, \cohSecond, \mutualCoh\}$. Without loss of generality, we assume that $\cohFirst \leq \cohSecond$. Let  $\sparsityFirst$ and $\sparsitySecond$ be nonnegative integers such that
	\begin{equation}
		\sparsityFirst + \sparsitySecond < \max\left\{\frac{2(1 + \cohSecond)}{\cohSecond + 2\coh_\text{max}+ \sqrt{\cohSecond^2 + \mutualCoh^2}}, \frac{1+\coh_\text{max}}{2\coh_\text{max}}\right\}.
		\label{result theorem split 1}
	\end{equation}
	Then, the solution $(\sigFirstOpt,\,\sigSecondOpt)$ to the convex program~$(\probSplit)$ satisfies
	\begin{equation}
		\|\sigOpt - \sig\|_2 \leq C_0\,\varepsilon + C_1\!\left(\sigma_{\sparsityFirst}(\analysisFirst\sigFirst) + \sigma_{\sparsitySecond}(\analysisSecond\sigSecond)\right),
		\label{split threshold}
	\end{equation}
	where $C_0, C_1 \geq 0$ are constants that do not depend on $\sigFirst$ and~$\sigSecond$ and where $\sigOpt = [{\sigFirstOpt}^T \  {\sigSecondOpt}^T]^T$.
	\label{thm split}
\end{thm}

Note that the quantities $\cohFirst$, $\cohSecond$, and $\coh_m$ characterize the interplay between the measurement matrix $\measMat$ and the sparsifying transforms $\analysisFirst$ and $\analysisSecond$.

As a corollary to our main result, we get the following statement for the problem $(\prob)$ considered in~\cite{Candes:2011p12}.
\begin{cor}
	Let $\meas = \measMat\sig + \noise$ with $\|\noise\|_2 \leq \varepsilon$ and let $\analysis \in \C^{\sizeAnalysis \times \dimSigHybrid}$ be a full-rank matrix. 
	Let $\sparsity$ be a nonnegative integer such that
	\begin{equation}
		\sparsity < \frac{1}{2}\!\left(1 + \frac{1}{\coh(\measMat\analysis^\dagger)}\right).
		\label{general recovery condition}
	\end{equation}
	Then, the solution $\sigOpt$ to the convex program~$(\prob)$ satisfies
	\begin{equation}
		\|\sigOpt - \sig\|_2 \leq C_0\,\varepsilon + C_1\sigma_k(\analysis\sig),
		\label{general threshold}
	\end{equation}
	where $C_0, C_1\geq 0$ are constants\footnote{Note that the constants $C_0$ and $C_1$ may take on different values at each occurrence.} that do not depend on $\sig$.
	\label{cor general}
\end{cor}

The proofs of Theorem~\ref{thm split} and Corollary~\ref{cor general} can be found in the Appendix.

We conclude by noting that D-RIP recovery guarantees for~$(\prob)$ were provided in~\cite{Candes:2011p12}. As is common in RIP-based recovery guarantees the restricted isometry constants are, in general, hard to compute. Moreover, the results in~\cite{Candes:2011p12} hinge on the assumption that $\analysis$ forms a Parseval frame, i.e., $\analysis^H\analysis = \idMat_\dimSig$; a corresponding extension to general $\analysis$ was provided in~\cite{Liu:2012p3466}. We finally note that it does not seem possible to infer the coherence-based threshold (\ref{general recovery condition}) from the D-RIP recovery guarantees in~\cite{Candes:2011p12,Liu:2012p3466}.

\section{Numerical Results}
\label{numericalResults}


We analyze an image-separation problem where we remove a fingerprint from a cartoon image. We corrupt the $512 \times 512$ greyscale cartoon image depicted in Fig.~\ref{cartoon} by adding a fingerprint\footnote{The fingerprint image is taken from {\scriptsize{\texttt{\url{http://commons.wikimedia.org/}}}}} and i.i.d. zero-mean Gaussian noise.

\newcommand{\image}{\mathbf{I}}
\newcommand{\finiteDiff}{\boldsymbol{\Delta}}
\newcommand{\vect}{\text{Vec}}

Cartoon images are constant apart from (a small number of) discontinuities and are thus sparse under the finite difference operator $\finiteDiff$ defined in~\cite{NAM:2011:INRIA-00602205:1}. 
Fingerprints are sparse under the application of a wave atom transform, $\mathbf{W}$, such as the redundancy~$2$ transform available in the WaveAtom toolbox\footnote{We used the WaveAtom toolbox from~{\scriptsize{\texttt{\url{http://www.waveatom.org/}}}}}~\cite{Demanet:2007p3635}. 
It is therefore sensible to perform separation by solving the problem $(\probSplitAnalysis)$ with $\analysisFirst = \finiteDiff$, $\analysisSecond = \mathbf{W}$, and $\measMat = \idMat_\dimSig$. For our simulation, we use a regularized version of $\finiteDiff$ and we employ the TFOCS solver\footnote{We used TFOCS from {\scriptsize{\texttt{\url{http://tfocs.stanford.edu/}}}}} from~\cite{tfocs}.

Fig.~\ref{cartoon restored} shows the corresponding recovered image. We can see that the restoration procedure gives visually satisfactory results.


\appendices

\section{Proofs}
\label{appendix proof thm}

For simplicity of exposition, we first present the proof of Corollary~\ref{cor general} and then describe the proof of Theorem~\ref{thm split}.

\subsection{Proof of Corollary~\ref{cor general}}


We define the vector $\kerVec = \sigOpt - \sig$, where $\sigOpt$ is the solution to $(\prob)$ and $\sig$ is the vector to be recovered. We furthermore set $\suppSig = \supp_\sparsity(\analysis\sig)$.

\subsubsection{Prerequisites}
Our proof relies partly on two important results developed earlier in \cite{Candes:2011p12}, \cite{Candes:2005p164} and summarized, for completeness, next.

\begin{lem}[Cone constraint \!\!\!\cite{Candes:2011p12},\! \cite{Candes:2005p164}]
\label{cone constraint lem}The vector $\analysis\kerVec$ obeys
\begin{equation}
	\|\analysis_{\suppSig^c}\kerVec\|_1 \leq \|\analysis_\suppSig\kerVec\|_1 + 2\|\analysis_{\suppSig^c}\sig\|_1,
	\label{cone constraint}
\end{equation}
where $\suppSig = \supp_\sparsity(\analysis\sig)$.
\end{lem}

\begin{IEEEproof}
Since $\sigOpt$ is the minimizer of $(\prob)$, the inequality $\|\analysis\sig\|_1 \geq \|\analysis\sigOpt\|_1$ holds.
Using $\analysis = \analysis_\suppSig + \analysis_{\suppSig^c}$ and $\sigOpt = \sig + \kerVec$, we obtain
\begin{subequations}
\begin{align}
	&\|\analysis_\suppSig\sig\|_1 + \|\analysis_{\suppSig^c}\sig\|_1 =  \|\analysis\sig\|_1 \notag\\
	&\quad \geq \|\analysis\sigOpt\|_1 =  \|\analysis_\suppSig\sig + \analysis_\suppSig\kerVec\|_1 + \|\analysis_{\suppSig^c}\sig + \analysis_{\suppSig^c}\kerVec\|_1\notag\\
	&\quad \geq \|\analysis_\suppSig\sig\|_1 - \|\analysis_\suppSig\kerVec\|_1 + \|\analysis_{\suppSig^c}\kerVec\|_1 - \|\analysis_{\suppSig^c}\sig\|_1. \notag
\end{align}
\end{subequations}
We retrieve (\ref{cone constraint}) by simple rearrangement of terms.
\end{IEEEproof}

\begin{lem}[Tube constraint \!\!\!\cite{Candes:2011p12},\! \cite{Candes:2005p164}]
The vector $\measMat\kerVec$ satisfies $\|\measMat\kerVec\|_2 \leq 2\varepsilon$.
\label{tube constraint}
\end{lem}
\begin{IEEEproof}
Since both $\sigOpt$ and $\sig$ are feasible (we recall that $\meas = \measMat\sig + \noise$ with $\|\noise\|_2 \leq \varepsilon$), we have the following
\begin{align}
	\|\measMat\kerVec\|_2 &= \|\measMat(\sigOpt - \sig)\|_2 \notag\\
	&\leq \|\measMat\sigOpt - \meas\|_2 + \|\meas - \measMat\sig\|_2 \leq 2\varepsilon, \notag
\end{align}
thus establishing the lemma.
\end{IEEEproof}

\nopagebreak

\subsubsection{Bounding the recovery error}
We want to bound $\|\kerVec\|_2$ from above. Since $\sigmaMin(\analysis) > 0$ by assumption ($\analysis$ is assumed to be full-rank), it follows from the Rayleigh-Ritz theorem~\cite[Th.~4.2.2]{HornJohnson} that
\begin{equation}
	\|\kerVec\|_2 \leq \frac{1}{\sigmaMin(\analysis)}\|\analysis\kerVec\|_2. \label{first upper bound proof 0}
\end{equation}

We now set $\maxCoeffSet = \supp_\sparsity(\analysis\kerVec)$. Clearly, we have for $\rowIdx \in \maxCoeffSet^c$,
\begin{equation}
	|[\analysis\kerVec]_\rowIdx| \leq \frac{\|\analysis_\maxCoeffSet\kerVec\|_1}{\sparsity} \notag.
\end{equation}
Using the same argument as in~\cite[Th.~3.1]{Cai:2010p142}, we obtain
\begin{align}
	\|\analysis_{\maxCoeffSet^c}\kerVec\|_2^2 = \sum_{\rowIdx\in\maxCoeffSet^c} |[\analysis\kerVec]_\rowIdx|^2 &\leq \sum_{\rowIdx\in\maxCoeffSet^c} |[\analysis\kerVec]_\rowIdx|\frac{\|\analysis_\maxCoeffSet\kerVec\|_1}{\sparsity} \notag\\
	&= \|\analysis_{\maxCoeffSet^c}\kerVec\|_1\frac{\|\analysis_\maxCoeffSet\kerVec\|_1}{\sparsity}. \label{first upper bound proof 1}
\end{align}
Since $\maxCoeffSet$ is the set of indices of the $\sparsity$ largest (in magnitude) coefficients of $\analysis\kerVec$ and since $\maxCoeffSet$ and $\suppSig$ both contain $\sparsity$ elements, we have $\|\analysis_\suppSig\kerVec\|_1 \leq \|\analysis_\maxCoeffSet\kerVec\|_1$ and  $\|\analysis_{\maxCoeffSet^c}\kerVec\|_1 \leq \|\analysis_{\suppSig^c}\kerVec\|_1$, which, combined with the cone constraint in Lemma~\ref{cone constraint lem}, yields
\begin{equation}
	\|\analysis_{\maxCoeffSet^c}\kerVec\|_1 \leq \|\analysis_\maxCoeffSet\kerVec\|_1 + 2\|\analysis_{\suppSig^c}\sig\|_1. \label{cone constraint with q}
\end{equation}
The inequality in (\ref{first upper bound proof 1}) then becomes
\begin{subequations}
\begin{align}
	\|\analysis_{\maxCoeffSet^c}\kerVec\|_2^2 &\leq \frac{\|\analysis_\maxCoeffSet\kerVec\|_1^2}{\sparsity} + 2\|\analysis_{\suppSig^c}\sig\|_1\frac{\|\analysis_\maxCoeffSet\kerVec\|_1}{\sparsity} \notag\\
								  &\leq \|\analysis_\maxCoeffSet\kerVec\|_2^2 + 2\|\analysis_{\suppSig^c}\sig\|_1\frac{\|\analysis_\maxCoeffSet\kerVec\|_2}{\sqrt{\sparsity}} \label{first upper bound proof 2}\\
								  &\leq 2\|\analysis_\maxCoeffSet\kerVec\|_2^2 + \frac{\|\analysis_{\suppSig^c}\sig\|_1^2}{\sparsity}, \label{first upper bound proof 3}
\end{align}
\end{subequations}
where (\ref{first upper bound proof 2}) follows from $\|\mathbf{u}\|_1 \leq \sqrt{\sparsity}\|\mathbf{u}\|_2$ for $\sparsity$-sparse\footnote{A vector is $\sparsity$-sparse if it has at most $\sparsity$ nonzero entries.} $\mathbf{u}$ and (\ref{first upper bound proof 3}) is a consequence of $2xy \leq x^2 + y^2$, for $x, y \in \R$.

It now follows that
\begin{subequations}
\begin{align}
	\|\analysis\kerVec\|_2 &= \sqrt{\|\analysis_\maxCoeffSet\kerVec\|_2^2 + \|\analysis_{\maxCoeffSet^c}\kerVec\|_2^2} \notag\\
						  &\leq \sqrt{3\|\analysis_\maxCoeffSet\kerVec\|_2^2 + \frac{\|\analysis_{\suppSig^c}\sig\|_1^2}{\sparsity}} \label{first upper bound proof 4}\\
						  &\leq \sqrt{3}\|\analysis_\maxCoeffSet\kerVec\|_2 + \frac{\|\analysis_{\suppSig^c}\sig\|_1}{\sqrt{\sparsity}}, \label{first upper bound proof 5}
\end{align}
\end{subequations}
where (\ref{first upper bound proof 4}) is a consequence of (\ref{first upper bound proof 3}) and (\ref{first upper bound proof 5}) results from $\sqrt{x^2 + y^2} \leq x + y$, for $x, y \geq 0$.

Combining (\ref{first upper bound proof 0}) and (\ref{first upper bound proof 5}) leads to
\begin{equation}
	\|\kerVec\|_2 \leq \frac{1}{\sigmaMin(\analysis)}\!\left(\sqrt{3}\|\analysis_\maxCoeffSet\kerVec\|_2 + \frac{\|\analysis_{\suppSig^c}\sig\|_1}{\sqrt{\sparsity}}\right). \label{upper bound h}
\end{equation}

\subsubsection{Bounding the term $\|\analysis_\maxCoeffSet\kerVec\|_2$ in (\ref{upper bound h})}

In the last step of the proof, we bound the term $\|\analysis_\maxCoeffSet\kerVec\|_2$ in (\ref{upper bound h}). To this end, we first bound $\|\measMat\analysis^\dagger\analysis_\maxCoeffSet\kerVec\|_2^2$, with  $\analysis^\dagger = (\analysis^H\analysis)^{-1}\analysis^H$, using
Ger\v{s}gorin's disc theorem~\cite[Th.~6.2.2]{HornJohnson}:
\begin{equation}
	\thetaMin\|\analysis_\maxCoeffSet\kerVec\|_2^2 \leq \|\measMat\analysis^\dagger\analysis_\maxCoeffSet\kerVec\|_2^2 \leq \thetaMax\|\analysis_\maxCoeffSet\kerVec\|_2^2 \label{general rip}
\end{equation}
where $\thetaMin \triangleq \minNorm^2 - \maxCorr(\sparsity - 1)$ and $\thetaMax \triangleq \maxNorm^2 + \maxCorr(\sparsity - 1)$
with
\begin{equation}
	\maxCorr = \underset{\rowIdx, \colIdx, \rowIdx \neq \colIdx}{\max}\,|[(\measMat\analysis^\dagger)^H\measMat\analysis^\dagger]_{\rowIdx,\colIdx}| \label{def maxcorr}
\end{equation}
and $\minNorm \triangleq \minNorm(\measMat\analysis^\dagger)$ and $\maxNorm \triangleq \maxNorm(\measMat\analysis^\dagger)$.

Using Lemma~\ref{tube constraint} and (\ref{general rip}) and following the same steps as in~\cite[Th.~2.1]{Cai:2010p141} and \cite[Th.~1]{Studer:2012p15}, we arrive at the following chain of inequalities:
\begin{subequations}
\begin{align}
	&\thetaMin\|\analysis_\maxCoeffSet\kerVec\|_2^2 \leq \|\measMat\analysis^\dagger\analysis_\maxCoeffSet\kerVec\|_2^2 = (\measMat\analysis^\dagger\analysis_\maxCoeffSet\kerVec)^H\measMat\analysis^\dagger\analysis_\maxCoeffSet\kerVec \notag\\
	&\!\!\!= (\measMat\kerVec)^H\measMat\analysis^\dagger\analysis_\maxCoeffSet\kerVec - (\measMat\analysis^\dagger\analysis_{\maxCoeffSet^c}\kerVec)^H\measMat\analysis^\dagger\analysis_\maxCoeffSet\kerVec \label{general proof 1}\\
	&\!\!\!\leq |(\measMat\kerVec)^H\measMat\analysis^\dagger\analysis_\maxCoeffSet\kerVec| + |(\analysis_{\maxCoeffSet^c}\kerVec)^H(\measMat\analysis^\dagger)^H\measMat\analysis^\dagger(\analysis_\maxCoeffSet\kerVec)| \notag\\
	&\!\!\!\leq \|\measMat\kerVec\|_2\|\measMat\analysis^\dagger\analysis_\maxCoeffSet\kerVec\|_2 \notag\\
	&\qquad\,\,\,+ \sum_{\rowIdx \in \maxCoeffSet^c,\colIdx \in \maxCoeffSet} |[(\measMat\analysis^\dagger)^H\measMat\analysis^\dagger]_{\rowIdx,\colIdx}||[\analysis\kerVec]_\rowIdx||[\analysis\kerVec]_\colIdx| \label{general proof 2}\\
	&\!\!\!\leq 2\varepsilon\sqrt{\thetaMax}\|\analysis_\maxCoeffSet\kerVec\|_2 + \maxCorr\|\analysis_\maxCoeffSet\kerVec\|_1\|\analysis_{\maxCoeffSet^c}\kerVec\|_1 \label{general proof 3} \\
	&\!\!\!\leq 2\varepsilon\sqrt{\thetaMax}\|\analysis_\maxCoeffSet\kerVec\|_2 + \maxCorr\|\analysis_\maxCoeffSet\kerVec\|_1\left(\|\analysis_\maxCoeffSet\kerVec\|_1 + 2\|\analysis_{\suppSig^c}\sig\|_1\right) \label{general proof 4}\\
	&\!\!\!\leq 2\varepsilon\sqrt{\thetaMax}\|\analysis_\maxCoeffSet\kerVec\|_2 + \maxCorr\sparsity\|\analysis_\maxCoeffSet\kerVec\|_2^2 \notag\\
	&\qquad\,\,\,+ 2\maxCorr\sqrt{\sparsity}\|\analysis_{\suppSig^c}\sig\|_1\|\analysis_\maxCoeffSet\kerVec\|_2, \label{general proof 5}
\end{align}
\end{subequations}
where (\ref{general proof 1}) follows from $\analysis_\maxCoeffSet\kerVec = \analysis\kerVec -\analysis_{\maxCoeffSet^c}\kerVec$ and $\analysis^\dagger\analysis = \idMat_\dimSig$, (\ref{general proof 2}) is a consequence of the Cauchy-Schwarz inequality, (\ref{general proof 3}) is obtained from (\ref{general rip}), Lemma~\ref{tube constraint}, and the definition of $\maxCorr$ in (\ref{def maxcorr}), (\ref{general proof 4}) results from (\ref{cone constraint with q}), and (\ref{general proof 5}) comes from $\|\mathbf{u}\|_1 \leq \sqrt{k}\|\mathbf{u}\|_2$, for $k$-sparse $\mathbf{u}$.

If $\kerVec \neq \zeroMat$, then $\|\analysis_\maxCoeffSet\kerVec\|_2 \neq 0$, since $\analysis$ is assumed to be full-rank and $\maxCoeffSet$ is the set of indices of the $\sparsity$ largest (in magnitude) coefficients of $\analysis\kerVec$, and therefore, the inequality between $\thetaMin\|\analysis_\maxCoeffSet\kerVec\|_2^2 $ and (\ref{general proof 5}) simplifies~to
\begin{equation}
	(\minNorm^2 - \maxCorr(2\sparsity - 1))\|\analysis_\maxCoeffSet\kerVec\|_2 \leq 2\varepsilon\sqrt{\thetaMax} + 2\maxCorr\sqrt{\sparsity}\|\analysis_{\suppSig^c}\sig\|_1. \notag
\end{equation}
This finally yields
\begin{equation}
	\|\analysis_\maxCoeffSet\kerVec\|_2 \leq \frac{2\varepsilon\sqrt{\thetaMax} + 2\maxCorr\sqrt{\sparsity}\|\analysis_{\suppSig^c}\sig\|_1}{\minNorm^2 - \maxCorr(2\sparsity - 1)}
	\label{upper bound phi h}
\end{equation}
provided that
\begin{equation}
	\minNorm^2 - \maxCorr(2\sparsity - 1) > 0. \notag
\end{equation}

\subsubsection{Recovery guarantee}
Using Definition~\ref{def coherence}, we get $\coh = \coh(\measMat\analysis^\dagger) = \maxCorr / \minNorm^2$.
Combining (\ref{upper bound h}) and (\ref{upper bound phi h}), we therefore conclude that for 
\begin{equation}
	\sparsity < \frac{1}{2}\!\left(1 + \frac{1}{\coh}\right) \label{eq eq}
\end{equation}
we have
\begin{equation}
	\|\sigOpt - \sig\|_2 = \|\kerVec\|_2  \leq C_0\,\varepsilon + C_1\|\analysis_{\suppSig^c}\sig\|_1 \notag
\end{equation}
with 
\begin{align}
	C_0 &= \frac{2\sqrt{3}}{\sigmaMin(\analysis)\minNorm}\frac{\sqrt{\frac{\maxNorm^2}{\minNorm^2}(1 + \coh(\sparsity - 1)})}{1 - \coh(2\sparsity - 1)} \notag\\
	C_1 & = \frac{1}{\sigmaMin(\analysis)}\!\left(\frac{2\coh\sqrt{3\sparsity}}{1 - \coh(2\sparsity - 1)} + \frac{1}{\sqrt{\sparsity}}\right). \notag
\end{align}

\subsection{Proof of Theorem~\ref{thm split}}

We start by transforming $(\probSplit)$ into the equivalent problem
\begin{equation}
	(\prob)\,\,\, \underset{\sigCompoundProb}{\minimize}\quad\|\analysisCompound\sigCompoundProb\|_1 \quad \st \quad \|\meas - \measMatCompound\sigCompoundProb\|_2 \leq \varepsilon \notag
\end{equation}
by amalgamating $\analysisFirst, \analysisSecond$ and $\measMatFirst, \measMatSecond$ into the matrices $\analysisCompound$ and $\measMatCompound$ as  follows:
\begin{align}
	\measMatCompound &= \begin{bmatrix}\measMatFirst & \measMatSecond\end{bmatrix} \in \C^{\measNb \times p} \label{measMat split}\\
	\analysisCompound &= \begin{bmatrix}\analysisFirst & \zeroMat_{\sizeAnalysis \times \dimSig} \\
						   \zeroMat_{\sizeAnalysis \times \dimSig} & \analysisSecond
			\end{bmatrix}
			\in \C^{2\sizeAnalysis\times2\dimSig}, \label{analysis split}
\end{align}
where $p = 2\dimSig$ in the analysis setting, $p = 2\sizeAnalysis$ in the synthesis setting, and $p = \dimSig + \sizeAnalysis$ in hybrid settings.
The corresponding measurement vector is $\meas = \measMatCompound\sigCompound + \noise$, where we set $\sigCompound = [\sigFirst^T\ \sigSecond^T]^T$.

A recovery condition for $(\probSplit)$ could now be obtained by simply inserting $\measMatCompound$ and $\analysisCompound$ in (\ref{measMat split}), (\ref{analysis split}) above into (\ref{general recovery condition}). In certain cases, we can, however, get a better (i.e., less restrictive) threshold following ideas similar to those reported in \cite{Studer:2012p15} and detailed next.

We define the vectors $\kerVecCompound_1 = \sigFirstOpt - \sigFirst$, $\kerVecCompound_2 = \sigSecondOpt - \sigSecond$, the sets $\maxCoeffSetFirst \triangleq \supp_{\sparsityFirst}(\analysisFirst\kerVecCompound_1)$, $\maxCoeffSetSecond \triangleq \sizeAnalysis + \supp_{\sparsitySecond}(\analysisSecond\kerVecCompound_2)$, and $\kerVecCompound = [\kerVecCompound_1^T\ \kerVecCompound_2^T]^T$, $\maxCoeffSet = \maxCoeffSetFirst \cup \maxCoeffSetSecond$, and set $\sparsity = \sparsityFirst + \sparsitySecond$.

We furthermore let, for $\ell = 1,2$,
$$\maxCorr_\ell = \underset{\rowIdx, \colIdx, \rowIdx \neq \colIdx}{\max}\,|[(\measMat_\ell\analysis_\ell^\dagger)^H\measMat_\ell\analysis_\ell^\dagger]_{\rowIdx,\colIdx}|$$
$$\mutualMaxCorr = \underset{\rowIdx, \colIdx}{\max}\,|[(\measMatFirst\analysisFirst^\dagger)^H\measMatSecond\analysisSecond^\dagger]_{\rowIdx,\colIdx}|.$$

With the definitions of $\maxCoeffSetFirst$ and $\maxCoeffSetSecond$, we have from~(\ref{general rip})
\begin{align}
	\|\measMatCompound\analysisCompound^\dagger\analysisCompound_\maxCoeffSet\kerVecCompound\|_2^2 &= \|\measMatCompound\analysisCompound^\dagger\analysisCompound_{\maxCoeffSetFirst}\kerVecCompound\|_2^2 + \|\measMatCompound\analysisCompound^\dagger\analysisCompound_{\maxCoeffSetSecond}\kerVecCompound\|_2^2 &\notag\\
	&\qquad+ 2(\measMatCompound\analysisCompound^\dagger\analysisCompound_{\maxCoeffSetFirst}\kerVecCompound)^H\measMatCompound\analysisCompound^\dagger\analysisCompound_{\maxCoeffSetSecond}\kerVecCompound. \label{mixed term}
\end{align}
The application of Ger\v{s}gorin's disc theorem~\cite{HornJohnson} gives
\begin{align}
	&\theta_{\text{min}, 1}\|\analysisCompound_{\maxCoeffSetFirst}\kerVecCompound\|_2^2 \leq \|\measMatCompound\analysisCompound^\dagger\analysisCompound_{\maxCoeffSetFirst}\kerVecCompound\|_2^2 \leq \theta_{\text{max}, 1}\|\analysisCompound_{\maxCoeffSetFirst}\kerVecCompound\|_2^2 \label{rip 1}\\
	&\theta_{\text{min}, 2}\|\analysisCompound_{\maxCoeffSetSecond}\kerVecCompound\|_2^2 \leq \|\measMatCompound\analysisCompound^\dagger\analysisCompound_{\maxCoeffSetSecond}\kerVecCompound\|_2^2 \leq \theta_{\text{max}, 2}\|\analysisCompound_{\maxCoeffSetSecond}\kerVecCompound\|_2^2\label{rip 2}
\end{align}
with $\theta_{\text{min}, \ell} \triangleq \minNorm^2(\measMat_\ell\analysisCompound_\ell^\dagger) - \maxCorr_\ell(\sparsity_\ell - 1)$ and $\theta_{\text{max}, \ell} \triangleq  \maxNorm^2(\measMat_\ell\analysisCompound_\ell^\dagger)  + \maxCorr_\ell(\sparsity_\ell - 1)$,  for $\ell = 1, 2$.

In addition, the last term in (\ref{mixed term}) can be bounded as
\begin{subequations}
\begin{align}
	&|(\measMatCompound\analysisCompound^\dagger\analysisCompound_{\maxCoeffSetFirst}\kerVecCompound)^H\measMatCompound\analysisCompound^\dagger\analysisCompound_{\maxCoeffSetSecond}\kerVecCompound| \notag\\
	&\quad\leq \sum_{\rowIdx \in \maxCoeffSetFirst, \colIdx \in \maxCoeffSetSecond} |[(\measMatCompound\analysisCompound^\dagger)^H\measMatCompound\analysisCompound^\dagger]_{\rowIdx,\colIdx}||[\analysisCompound\kerVecCompound]_\rowIdx||[\analysisCompound\kerVecCompound]_\colIdx| \notag\\
	&\quad\leq \mutualMaxCorr\|\analysisCompound_{\maxCoeffSetFirst}\kerVecCompound\|_1\|\analysisCompound_{\maxCoeffSetSecond}\kerVecCompound\|_1 \label{mixed 1}\\
	&\quad\leq \mutualMaxCorr\sqrt{\sparsityFirst\sparsitySecond}\|\analysisCompound_{\maxCoeffSetFirst}\kerVecCompound\|_2\|\analysisCompound_{\maxCoeffSetSecond}\kerVecCompound\|_2 \label{mixed 2}\\
	&\quad\leq \frac{\mutualMaxCorr}{2}\sqrt{\sparsityFirst\sparsitySecond}\left(\|\analysisCompound_{\maxCoeffSetFirst}\kerVecCompound\|_2^2 + \|\analysisCompound_{\maxCoeffSetSecond}\kerVecCompound\|_2^2\right)\label{mixed 3}\\
	&\quad\leq \frac{\mutualMaxCorr}{2}\sqrt{\sparsityFirst\sparsitySecond}\|\analysisCompound_\maxCoeffSet\kerVecCompound\|_2^2, \label{rip mixed}
\end{align}
\end{subequations}
where (\ref{mixed 1}) follows from the definition of $\mutualMaxCorr$, (\ref{mixed 2}) results from $\|\mathbf{u}\|_1 \leq \sqrt{k}\|\mathbf{u}\|_2$, for $k$-sparse $\mathbf{u}$, and (\ref{mixed 3}) is a consequence of the arithmetic-mean geometric-mean inequality.

Combining (\ref{rip 1}), (\ref{rip 2}), and (\ref{rip mixed}) gives
\begin{equation}
	\thetaMin\|\analysisCompound_\maxCoeffSet\kerVecCompound\|_2^2 \leq \|\measMatCompound\analysisCompound^\dagger\analysisCompound_\maxCoeffSet\kerVecCompound\|_2^2 \leq \thetaMax\|\analysisCompound_\maxCoeffSet\kerVecCompound\|_2^2, \notag
\end{equation}
where $\thetaMin \triangleq \minNorm^2 - f(\sparsityFirst, \sparsitySecond)$, $\thetaMax \triangleq \maxNorm^2 + f(\sparsityFirst, \sparsitySecond)$, $\minNorm \triangleq \minNorm(\measMat_1\analysis_1^\dagger, \measMat_2\analysis_2^\dagger)$, $\maxNorm \triangleq \maxNorm(\measMat_1\analysis_1^\dagger,\measMat_2\analysis_2^\dagger)$, and 
\begin{equation}
	f(\sparsityFirst, \sparsitySecond) \triangleq \max\{\maxCorrFirst(\sparsityFirst - 1), \maxCorrSecond(\sparsitySecond - 1)\} + \mutualMaxCorr\sqrt{\sparsityFirst\sparsitySecond}. \notag
\end{equation}
Using the same steps as in (\ref{general proof 1})-(\ref{general proof 5}), we get
\begin{equation}
	g(\sparsityFirst, \sparsitySecond)\|\analysisCompound_\maxCoeffSet\kerVecCompound\|_2 \leq 2\varepsilon\sqrt{\thetaMax} + 2\maxCorr\sqrt{\sparsity}\|\analysisCompound_{\suppSig^c}\sigCompound\|_1, \notag
\end{equation}
where $g(\sparsityFirst, \sparsitySecond) \triangleq \minNorm^2 - f(\sparsityFirst, \sparsitySecond) - \maxCorr\sparsity$.	

Next, we bound $g(\sparsityFirst, \sparsitySecond)$ from below by a function of $\sparsity = \sparsityFirst + \sparsitySecond$. This can be done, e.g., by looking for the minimum~\cite{Studer:2012p15}
\begin{equation}
	\hat{g}(\sparsity) \triangleq \underset{\sparsityFirst\colon 0 \leq \sparsityFirst \leq \sparsity}{\min}\,g(\sparsityFirst, \sparsity - \sparsityFirst) \label{max g 1}
\end{equation}
or equivalently
\begin{equation}
	\hat{g}(\sparsity) \triangleq \underset{\sparsitySecond\colon 0 \leq \sparsitySecond \leq \sparsity}{\min}\,g(\sparsity - \sparsitySecond, \sparsitySecond). \label{max g 2}
\end{equation}
To find $\hat{g}(\sparsity)$ in (\ref{max g 1}) or in (\ref{max g 2}), we need to distinguish between two cases: 

\textbullet~\underline{Case 1:} $\maxCorrFirst(\sparsityFirst - 1) \leq \maxCorrSecond(\sparsitySecond - 1)$\\
In this case, we get
\begin{equation}
	g(\sparsity - \sparsitySecond, \sparsitySecond) = \minNorm^2 - \maxCorrSecond(\sparsitySecond - 1) - \mutualMaxCorr\sqrt{\sparsitySecond(\sparsity - \sparsitySecond)} - \maxCorr\sparsity.\notag
\end{equation}
A straightforward calculation reveals that the minimum of $g$ is achieved at
\begin{equation}
	\sparsitySecond = \frac{\sparsity}{2}\!\left(1 + \frac{\maxCorrSecond}{\sqrt{\maxCorrSecond^2 + \mutualMaxCorr^2}}\right), \notag
\end{equation}
resulting in
\begin{equation}
	\hat{g}(\sparsity) = \minNorm^2 - \frac{1}{2}\!\left(\maxCorrSecond(\sparsity - 2) + \sparsity\sqrt{\maxCorrSecond^2 + \mutualMaxCorr^2}\right) - \maxCorr\sparsity. \notag
\end{equation}
If $\hat{g}(\sparsity) > 0$, then we have
\begin{equation}
	\|\sigCompoundOpt - \sigCompound\|_2 = \|\kerVecCompound\|_2 \leq C_0\,\varepsilon + C_1\|\analysisCompound_{\suppSig^c}\sigCompound\|_1
\end{equation}
where
$$
	C_0 = \frac{2\sqrt{3}}{\sigmaMin(\analysisCompound)\hat{g}(\sparsity)}
$$
and
$$
	C_1 = \frac{1}{\sigmaMin(\analysisCompound)}\!\left(\frac{2\maxCorr\sqrt{3k}}{\hat{g}(\sparsity)} + \frac{1}{\sqrt{\sparsity}}\right).
$$
Setting $\hat{g}(\sparsity) > 0$ amounts to imposing
\begin{equation}
	\sparsity < \frac{2\left(1+ \cohSecond\right)}{\cohSecond + 2\coh_\text{max} + \sqrt{\cohSecond^2 + \mutualCoh^2}}, \label{ineq1}
\end{equation}
where we used Definitions~\ref{def coherence} and \ref{def mutual coherence} to get a threshold depending on the coherence parameters only. 

\textbullet~\underline{Case 2:} $\maxCorrSecond(\sparsitySecond - 1) \leq \maxCorrFirst(\sparsityFirst - 1)$\\
Similarly to Case~1, we get
\begin{equation}
	\hat{g}(\sparsity) = \minNorm^2 - \frac{1}{2}\!\left(\maxCorrFirst(\sparsity - 2) + \sparsity\sqrt{\maxCorrFirst^2 + \mutualMaxCorr^2}\right) - \maxCorr\sparsity. \notag
\end{equation}
If $\hat{g}(\sparsity) > 0$, we must have 
\begin{equation}
	\sparsity < \frac{2\left(1+ \cohFirst\right)}{\cohFirst + 2\coh_\text{max} + \sqrt{\cohFirst^2 + \mutualCoh^2}}. \label{ineq2}
\end{equation}

Since $\cohFirst \leq \cohSecond$, by assumption, the inequality in~(\ref{ineq2}) is tighter than the one in~(\ref{ineq1}). We complete the proof by combining the thresholds in (\ref{eq eq}) and (\ref{ineq1}) to get (\ref{result theorem split 1}).

\renewcommand{\baselinestretch}{0.983}

\bibliographystyle{IEEEtran} 
\bibliography{bibSplitAnalysisAndSynthesis.bib,bibSplitAnalysisAndSynthesisBook.bib}

\end{document}